\documentclass[11pt]{revtex4}

\usepackage{amssymb,hyperref}
\usepackage{times,subfigure}

\newcommand{\nc}{\newcommand}
\nc{\ox}{\otimes}

\usepackage{graphicx}
\usepackage{bm}
\usepackage{amsmath}
\usepackage{amsthm}

\newcommand{\ket}[1]{|#1\rangle}

\newcommand{\bra}[1]{\langle#1|}

\newcommand{\proj}[1]{\ket{#1}\bra{#1}}

\newcommand{\beq}{\begin{equation}}
\newcommand{\eeq}{\end{equation}}

\newcommand{\Tr}{{\rm Tr}}

\newcommand{\cT}{\mathcal{T}}

\begin{document}

\title{The problem with the geometric discord}
\begin{abstract}
We argue that the geometric discord introduced in [B. Daki{\'c}, V. Vedral, and C. Brukner, Phys. Rev. Lett. 105, 190502 (2010)] is not a good measure for the quantumness of correlations, as it can increase even under trivial local reversible operations of the party whose classicality/non-classicality is not tested. On the other hand it is known that the standard, mutual-information based discord does not suffer this problem; a simplified proof of such a fact is given.
\end{abstract}

\author{M. Piani}
\affiliation{Institute for Quantum Computing and Department of Physics and Astronomy, University of Waterloo, Waterloo ON
N2L 3G1, Canada}

\maketitle
The geometric measure of quantum discord was introduced in~\cite{PhysRevLett.105.190502} as a parameter of the quantumness of correlations. In particular it is meant to quantify the distance---in Hilbert-Schmidt norm---of a bipartite quantum state from the set of classical-quantum states $\rho^{CQ}_{AB}=\sum_i p_i \proj{i}_A\otimes \rho^i_B$; as such it is asymmetric with respect to the two subsystems $A$ and $B$. Its definition (up to an irrelevant factor) is
\beq
\label{eq:geomdiscord}
D_G(B|A)_{\rho_{AB}}:=\inf_{\Pi_A}\|\rho_{AB}-\Pi_A (\rho_{AB})\|_2^2,
\eeq
with $\|X\|_2=\sqrt{\Tr(X^\dagger X)}$ the Hilbert-Schmidt norm, and the infimum---for finite-dimensional $A$, a minimum---running over complete von Neumann projections on $A$, i.e. $\Pi_A[X]=\sum_i\proj{i}X\proj{i}$ for some orthonormal basis $\{\ket{i}\}$ of $A$. The geometric measure has found widespread application because of its ease of use, in particular when $A$ is a qubit. It has been linked to the performance of remote state preparation~\cite{dakic2012quantum,tufarelli2012quantumness} and has attracted interest in its direct experimental quantification~\cite{1751-8121-45-11-115308,passante2012measuring,PhysRevLett.108.150403}. While it might be that in certain cases the geometric discord is a useful parameter of the quantumness of correlations, we will point out that arguably it cannot be anything more than that. Indeed, we will see that it can change arbitrarily and reversibly through actions of Bob (the unmeasured party in \eqref{eq:geomdiscord}). As such, it is hard to imagine that it might have any deep meaning---e.g., in an information-theoretic sense---or any fundamental operational interpretation~\footnote{It might still happen that within some restricted framework, like that of [10] there is a connection with some specific task; here we refer to something more abstract and general operational meaning.}. With this in mind, with this note we would like to draw the attention of the community on the potential risk of using the geometric discord as a basic quantifier of the quantumness of the correlations and in the analysis of the role of quantum correlations in fundamental tasks.

It is well known that measures of the quantumness of correlations~\cite{modi2011quantum}, contrary to entanglement measures~\cite{RevModPhys.81.865}, can increase under local actions of the parties. This is in particular true for the original discord measure defined in~\cite{zurek2000,PhysRevLett.88.017901}:
\beq
\label{eq:discord}
D(B|A)_{\rho_{AB}}=\inf_{\Pi_A}[I(A:B)_{\rho_{AB}}-I(A:B)_{\Pi_A (\rho_{AB})}],
\eeq
with $I(A:B)_{\tau_{AB}}:=S(\tau_A)+S(\tau_B)-S(\tau_AB)$ the mutual information and $S(\xi)=-\Tr(\xi\log_2\xi)$ the von Neumann entropy. Both the discord $D$ and the geometric discord $D_G$ vanish only for classical-quantum states. That means, for example, that an operation on $A$ can readily create discord: for example, for a channel (a completely positive trace-preserving map) $\Lambda$ acting as $\Lambda[\proj{0}]=\proj{0}$, $\Lambda[\proj{1}]=\proj{+}$, with $\ket{+}=(\ket{0}+\ket{1})/\sqrt{2}$, one has that the classical-classical---hence with zero discord---state $(\proj{0}\otimes\proj{0}+\proj{1}\otimes\proj{1})/2$ is mapped into the state $(\proj{0}\otimes\proj{0}+\proj{+}\otimes\proj{1})/2$ with non-zero discord by the action of $\Lambda_A$. While this fact might already be considered bothersome by some, it is not totally unreasonable: the creation of quantumness is done at the price of some loss of total correlations, as measured, for example, by mutual information, and it might be interpreted as the impossibility of treating the remaining correlations as fully classical from an information theoretic point of view (see, e.g., \cite{PhysRevLett.100.090502}). Moreover, the creation of quantumness takes place via an action on the system whose classicality is tested in the definition of the discord quantities. In particular, in~\cite{PhysRevLett.106.160401} (see \cite{PhysRevLett.106.220403} and \cite{PhysRevA.85.040301} for related results) it was proved that the discord $D$ cannot increase under actions of $B$. Here we give an alternative and simple proof of the same fact that is based solely on the monotonicity mutual information and may be of independent interest; the proof applies to any discord-like quantitiy---not necessarily meant to capture the quantumness of correlations---of the form
\beq
\label{eq:gendiscord}
D_\cT(B|A)_{\rho_{AB}}:=\inf_{\Lambda_A\in \cT}[I(A:B)_{\rho_{AB}}-I(A:B)_{\Lambda_A (\rho_{AB})}],
\eeq
where the infimum is over some class $\cT$ of channels $\Lambda_A$ on $A$. If such a class $\cT$ is that of complete projective measurements, one recovers the discord $D$ of Eq.~\eqref{eq:discord}; considering instead arbitrary measurements, i.e., $\Lambda[X]=\sum_i \Tr(M_iX) \proj{i}$, with $\{M_i\}_i$ a POVM and $\{\ket{i}\}$ orthonormal states~\footnote{The orthonormal states may span a much larger space than that of $A$, depending on the number of outcomes of the measurement.}, one obtains the other---POVM-based, rather than projection-based---standard version of quantum discord.

Our proof is based on rewriting the right-hand side of \eqref{eq:gendiscord} as
\begin{align*}
 &I(A:B)_{\rho_{AB}}-I(A:B)_{\Lambda_A (\rho_{AB})}\\
=\, &I(A'C:B)_{\rho_{A'BC}}-I(A':B)_{\rho_{A'BC}}\\
=\, &I(B:C|A')_{\rho_{A'BC}}.
\end{align*}
Here, we have used the fact that any channel from $A$ to $A'$ can be written as an isometry $V$ from $A$ to a composite system $A'C$ followed by the discarding of $C$, and we have made used of the definition $\rho_{A'BC}=V \rho_{AB} V^\dagger$. Thus, the first equality is due to the fact mutual information is invariant under local isometries and to the fact that $\Lambda_A[\rho_{AB}]=\Tr_C[\rho_{A'BC}]$. The second equality is simply the definition of the conditional mutual information $I(B:C|A'):=I(A'C:B) - I(A':B)$. The claim then follows form the monotonicity of conditional mutual information under channels on $B$~\footnote{Such monotonicity is simply the monotonicity of mutual information, since $I(B:C|A')$ can be rewritten as $I(B:C|A')=I(A':BC) - I(A':C)$, with $B$ appearing only in the first term.}.

The monotonicity of $D$ under operations on $B$ is comforting, since the definitions \eqref{eq:geomdiscord} and \eqref{eq:discord} are meant to capture the quantumness of correlations as due to the quantumness of the subsystem $A$. The problem with the geometric discord $D_G$ is that it does not have the just mentioned properties: it can increase under the action of the unmeasured party, and at no cost for total correlations, actually in a fully reversible way. At a more technical level, the source of the problem can be identified in the fact that the geometric discord $D_G$ of Eq.~\eqref{eq:geomdiscord} is based on a norm---the Hilbert-Schmidt norm---that is not monotonic under quantum evolutions (the application of channels), as pointed out, for example~\footnote{Ref.~\cite{Ozawa2000158} was motivated exactly by the need to clarify that the Hilbert-Schmidt distance was not an appropriate choice as the basis for the construction of a good distance-based entanglement measure.}, in Ref.~\cite{Ozawa2000158}. 
In this note we provide a simple case where monotonicity is violated, and use it to question the general validity of the geometric discord as a conceptually meaningful (rather than useful) parameter of quantumness.

Consider the simple channel $\Gamma^\sigma:X\rightarrow X\otimes \sigma$, i.e. the channel that introduced a noisy ancillary state. Under such an operation
\[
\|X\|_2\rightarrow \|\Gamma^\sigma X\|_2=\|X\|_2\|\sigma\|_2=\|X\|_2\sqrt{\Tr(\sigma^2)},
\]
since the Hilbert-Schmidt norm is multiplicative on tensor products. It is then easy to see that
\[
D_G(B|A)_{\Gamma^\sigma_B(\rho_{AB})}=D_G(B|A)_{\rho_{AB}}{\Tr(\sigma^2)},
\]
since the optimization on the projective measurement on $A$ is unaffected by the presence of factorized ancillary state on $B$.
Thus, adding or removing a factorized local ancilla---a local and reversible operation---adds or removes a factor equal to the purity of the ancillary state. Notice that one can even imagine the ancilla as always present, with only its state modified by $\Gamma^\sigma$. In particular, making the state of the uncorrelated ancilla purer---e.g., by just discarding the ancilla and preparing a new one in a purer state---increases the geometric discord.

A possible fix to prevent the geometric measure from increasing under local operations on $B$ is to trivially redefine it, for example as
\beq
\tilde{D}_G(B|A)_{\rho_{AB}}:=\sup_{\Lambda_B}{D}_G(B|A)_{\Lambda_B(\rho_{AB})},
\eeq
where the supremum is over channels on $B$ (not necessarily with output dimension equal to the input dimension). While this fixes by definition the problem of the increase of the measure under operations on $B$, it makes the (modified) geometric discord  in principle much more difficult to calculate, making the advantage of using a simple-to-calculate parameter of non-classicality disappear. Also, since  $\tilde{D}_G(B|A)$ would still be based on the non-monotonous Hilbert-Schmidt distance, it is to be expected that $\tilde{D}_G(B|A)$ could still present some unwanted issues from an operational---besides from a mathematical---point of view.

We conclude that the geometric discord based on the geometry induced by the Hilbert-Schmidt norm is arguably not the best conceptual and operational choice to quantify the quantumness of correlations, even if in some case it might be an interesting parameter to consider~\cite{dakic2012quantum,tufarelli2012quantumness}.

After completion of this note, it was pointed out to us that the observation that the geometric discord is not monotonic under operations on the unmeasured side was already made in \cite{hu2012quantum}, and further commented upon in Ref.~\cite{tufarelli2012quantumness}. In~\cite{hu2012quantum} a specific one-parameter example of such an occurence is given. We believe that the construction in this note emphasizes even more strongly the undesirable features of the geometric discord. It is worth pointing out that in \cite{hu2012quantum} the natural requirement that a one-sided measure of quantumness based on the test of the quantumness of $A$ should not increase under channels on $B$ is stressed and imposed as a prerequisite for a good quantumness measure. All in all, we believe it is still worth dragging more focused attention on the issue, so that steps can be taken by the community towards a critical analysis, definition, and use of quantumness measures. 

\emph{Acknowledgements.} We thank G. Adesso for discussions and for pointing out relevant and related points raised in Refs.~\cite{hu2012quantum} and \cite{tufarelli2012quantumness}. This note was for the large part completed during a visit to the National Quantum Information Centre (KCIK) in Gda\'nsk, whose hospitality is gratefully acknowledged.  This work has been supported by CIFAR and NSERC.


\end{document}